\documentclass[twocolumn]{svjour3}
\usepackage{graphicx}
\usepackage{bm}
\usepackage{ulem}
\usepackage{epstopdf}
\usepackage{epsfig}

\begin{document}

\title{Partial wave analysis of $K^+N$ scattering
and the possibility of pentaquark $\Theta^+(1540)$}

\author{Kook-Jin Kong \and Byung-Geel Yu}
\institute{Kook-Jin Kong \and Byung-Geel Yu
            \at Research Institute of Basic Science, Korea Aerospace
               University, Goyang, 10540, Korea
          \and
            \email{bgyu@kau.ac.kr}
            }

\date{\today}

\maketitle

\begin{abstract}
Elastic $K^+ n\to K^+ n$ and charge exchange $K^+n\to K^0p$ reactions
at low momenta are investigated using the partial wave analysis.
Isospin relation gives constrains for the partial $s$- and $p$-waves
among elastic $K^+p$, $K^+n$ and charge exchange $K^+n\to K^0p$ amplitudes.
Two sets of the phase shift for these partial waves in the isoscalar channel
are obtained from the fit of experimental data on total
and differential cross sections. Polarization observable leads to a good criterion
to decide which set is valid.
Differential cross section data near $P_{\rm Lab}=434$ MeV/$c$
suggests a possibility of a resonance, e.g., the exotic $\Theta^+(1540)$ baryon
around $\sqrt{s}\simeq$ 1.54GeV.
\end{abstract}


\section{Introduction}

According to the conventional quark model, a  meson consists of
$\{q\bar{q}\}$ and a baryon of $\{qqq\}$ or $\{\bar{q}\bar{q}\bar{q}\}$.
However, other states called the tetraquark  $\{q\bar{q}q\bar{q}\}$ and
pentaquark $\{qqqq\bar{q}\}$ may also exist \cite{lhcb}.
In search of the pentaquark state, the so
called $\Theta^+(1540)$ of the $\{uudd\bar{s}\}$ configuration,
T. Nakano {\it et\ al}.  claimed to
observe the exotic baryon resonance in the reaction $\gamma n\to K^+K^-n$
on the $^{12}C$ target at $1.54\pm0.01$ GeV with the width
less that 25 MeV \cite{nakano}, though more rigorous verification awaits since then.
A recent experiment using photon beam reported
a narrow peak at the missing mass around 1.54 GeV which could be the
$\Theta^+(1540)$ state in the reaction $\gamma p\to pK_S K_L$
\cite{amaryan}.
As the pentaquark configuration needs the $K^+\{u\bar{s}\}$ and neutron $\{udd\}$,
the elastic $K^+n\to K^+n$ and charge exchange $K^+n\to K^0p$ reactions
have long been the candidates of finding the $\Theta^+(1540)$ predicted at mass
$M \approx 1.53$ GeV/$c^2$ with the width $\Gamma < 15$ MeV \cite{diakonov,praszalowicz}.
The correction for the width was later suggested in Refs. \cite{jaffee,weigel}.
More recently, a theoretical study of the reaction $K^+d\to K^0p(p)$ predicted the
$\Theta^+(1540)$ peak most probable at $P_{\rm Lab}\approx0.4$ GeV/$c$ \cite{taka}.
In this regard, the $K^+N$ reaction \cite{arndt,azimov,hyslop} is intuitive
to investigate the possibility of finding the exotic baryon, $\Theta^+(1540)$.

Beyond the resonance region up to tens of GeV in the $K^+N$ reaction,
the peripheral scattering via the $t$-channel meson exchange
becomes dominant, and the reaction mechanism
is found to be governed by the tensor meson $f_2$ and
Pomeron exchanges in the isoscalar channel \cite{yu-kong-kn}.
In the low momentum region below the kaon laboratory momentum
$P_{\rm Lab}\leq 800$ MeV/$c$,
the meson exchange alone is not appropriate to
describe the differential and total cross-sections.
Therefore, we try to understand the $K^+N$ reaction
by using the partial wave analysis from threshold up to
$P_{\rm Lab}\approx800$ MeV/$c$
with our particular interest in the region
around $P_{\rm Lab}=434$ MeV/$c$ where the pentaquark $\Theta^+(1540)$ is
expected to exist.

In previous works there is a significant disagreement
between theory and experiment in the reaction cross sections near
$P_{\rm Lab}\approx434$ MeV/$c$.
A. Sibirtsev {\it et\ al}. \cite{Sibirtsev} calculated the $K^+d$ cross
section using the single scattering impulse approximation to find a large discrepancy
with data around $P_{\rm Lab}=434$ MeV/c in the case of the $K^+n$ elastic reaction.
Furthermore, K. Aoki {\it et\ al}. \cite{aoki} investigated the $K^+N$
cross section by employing the wave function renormalization method and
obtained the result inconsistent with experiment near $P_{\rm Lab}=434$ MeV/$c$.
All these numerical consequences are interesting in the sense that they require
a new approach to the region where the $\Theta^+(1540)$ is predicted to exist.

In this paper we will first work with the isovector amplitude from the
low momentum elastic $K^+p$ reaction, in which case only the $s$-wave is
considered from the isotropy of the reaction except for the Coulomb repulsion
at very forward angles.
After doing this, we will extract the isoscalar amplitudes from the elastic
$K^+n\to K^+n$ and charge exchange $K^+n\to K^0p$ reactions to investigate
whether the isoscalar exotic $\Theta^+(1540)$ exists in the expected region.
Though starting from a simple fit of parameters for the
$s$-wave phase shift for the $K^+ p$ elastic reaction,
the current approach is of value to describe the reaction cross
sections for the three channels with the phase shift of partial waves
in a unified way.

The paper is organized as follows. In Sec. II, we introduce
the reaction mechanism for the $K^+N$ reaction at low momenta in terms of
partial waves. Sec. III devotes to a discussion of numerical consequences
in total and differential cross sections including polarization observables
in comparison with experimental data. In Sec. IV,
based on our findings in the present analysis a perspective on the possibility
of finding the pentaquark $\Theta^+(1540)$ is given.

\section{Partial wave analysis for $K^+N$ reactions}

The $K^+N$ reaction consists of following three channels
\begin{eqnarray}
&&K^+p\to K^+p\,,\label{k+p}\\
&&K^+n\to K^0p\,,\label{k+ncex}\\
&&K^+n\to K^+n\,.\label{k+n}
\end{eqnarray}

Since the isospin of kaon is 1/2 in common with nucleon,
the elastic channel $K^+p\to K^+p$ is composed of the amplitude of
isospin $I=1$, and other two are the mixtures of isospin $I=1$ and 0.
Therefore, they are expressed as the sum of the
isoscalar($I=0$) and isovector($I=1$) amplitudes which are given by,
\begin{eqnarray}
&&{\cal M}(K^+p\to K^+p)={\cal M}^{(1)}+{\cal M}_C\,,\label{k+p-amp}\\
&&{\cal M}(K^+n\to K^0p)={1\over2}\left({\cal M}^{(1)}-{\cal
M}^{(0)}\right)\label{k+ncex-amp}\,,\\
&&{\cal M}(K^+n\to K^+n)={1\over2}\left({\cal M}^{(1)}+{\cal M}^{(0)}\right)
\label{k+n-amp},
\end{eqnarray}
where ${\cal M}^{(0)}\, ({\cal M}^{(1)})$ is the isoscalar (isovector) component of
the  reaction amplitude
and ${\cal M}_C$ is the Coulomb amplitude due to the repulsive interaction between
$K^+$ and proton \cite{aoki}.
Hence, the following relation holds for these three channels,
\begin{eqnarray}\label{channel-2}
&&{\cal M}(K^+n\to K^0p)\nonumber\\&&\hspace{1cm}={\cal M}(K^+p\to
K^+p)-{\cal M}(K^+n\to K^+n).
\end{eqnarray}

In practice, the experimental data on kaon scattering off a neutron target
are obtained from the scattering off a deuteron target
$K^+d \to K^+n(p)$ with the spectator proton \cite{damerell,giacomelli73}.
Therefore, the formula relevant should be modified to take into account the deuteron
form factors $I_0$ and $J_0$ \cite{hashimoto}. However, as the $I_0$'s are almost
1 except for the backward region and $J_0$'s are almost 0 except for the
forward region, these form factors can be ignored in the calculation.

Since the elastic channel $K^+p\to K^+p$ is of pure isovector,
we can easily find the isovector amplitude ${\cal M}^{(1)}$ in the
experimental data below 800 MeV/$c$.  Then, the isoscalar amplitude
${\cal M}^{(0)}$ is determined from the isospin relation for the $K^+n\to K^0p$ and
$K^+n\to K^+n$ reactions in Eq. (\ref{channel-2}) above.

\subsection{Isovector amplitude}

In the  elastic $K^+p\to K^+p$ scattering below $P_{\rm Lab}\approx 800$ MeV/$c$,
the total cross section is almost constant in the region.
Apart from the Coulomb repulsion, the flatness of the shape is due to the repulsive
hadronic interaction between $K^+$ and nucleon, which gives a hint at
the phase shift.
In the differential cross section, the angular dependence is
isotropic excluding the sharp peaks at very forward angles
due to the Coulomb repulsion.
To implement such an isotropy in the differential cross section,
a partial $s$-wave is considered with the phase shift.
Denoting it by the symbol $S_{11}$, the isovector amplitude for the $s$-wave
is written as
\begin{eqnarray}\label{phase}
S_{11}={1\over2ik}
\left(\eta^1_{0+} e^{2i\delta^1_{0+}}-1\right) 
\end{eqnarray}
with the inelasticity $\eta^1_{0+}=1$ for simplicity.

The phase shift of $S_{11}$ is obtained as a linear function of
the incident kaon momentum $k$ in the center of mass frame, i.e.,
\begin{eqnarray}
&&\delta^1_{0+}(k)=a_0+b_0 k \label{goldhaber}
\end{eqnarray}
with the coefficients $a_0=3$ and $b_0=-107$ GeV$^{-1}$ fixed
to the differential cross section data \cite{goldhaber,cameron}.
The phase shift is negative (see Fig. \ref{fig5} below) and consistent
with the repulsive hadron interaction between $K^+$ and proton.
Our fit in Eq. (\ref{goldhaber}) is almost the same as that of
Goldhaber \cite{goldhaber}, and hence, the total amplitude is
given by
\begin{eqnarray}\label{s11}
{\cal M}(K^+p\to K^+p) = S_{11}+{\cal M}_C
\end{eqnarray}
with the Coulomb interaction term ${\cal M}_C$ discussed in detail in Ref. \cite{aoki}.

\subsection{Isoscalar amplitude}

Now that we are dealing with the low momentum reaction below 800 MeV/$c$,
it is good to consider the $s$ and $p$-waves for the isoscalar amplitude.
Similar to the isovector case, the isoscalar $s$-wave amplitude is denoted as
\begin{eqnarray}\label{phase}
S_{01}={1\over2ik}
\left(\eta^0_{0+} e^{2i\delta^0_{0+}}-1\right),
\end{eqnarray}
and the partial $p$-waves are further constructed as,
\begin{eqnarray}
&&P_{01}=f^0_{1-}\cos\theta-i\vec{\sigma}\cdot\hat{n}f^0_{1-}\sin\theta,\\
&&P_{03}=2f^0_{1+}\cos\theta+i\vec{\sigma}\cdot\hat{n}f^0_{1+}\sin\theta,
\end{eqnarray}
where
\begin{eqnarray}\label{p1}
f^0_{1\pm}={1\over2ik}\left(\eta^0_{1\pm} e^{2i\delta^0_{1\pm}}-1\right)
\end{eqnarray}
with $\eta^0_{0+}=1$ and $\eta^0_{1\pm}=1$ for simplicity.

Thus, from the isospin relations in Eqs. (\ref{k+p-amp}),
(\ref{k+ncex-amp}) and (\ref{k+n-amp}) above,
the scattering amplitudes for $K^+n \to K^+n$ and
$K^+n \to K^0p$ reactions are written in terms of these $s$ and $p$ waves,
\begin{eqnarray}
&&{\cal M}(K^+n \to K^+n)={1\over2} \left(S_{11}+S_{01}+P_{01}+P_{03}\right)\,,
\label{phase-1}\\
&&{\cal M}(K^+n \to K^0p)={1\over2} \left(S_{11}-S_{01}-P_{01}-P_{03}\right)\,,
\label{phase-2}
\end{eqnarray}
which satisfies Eq.  (\ref{channel-2}).

For those elastic and charge exchange $K^+n$ reactions above, there are two sets of data
on the differential cross section measured by C. J. S. Damerell {\it et al}.
\cite{damerell} as presented in Fig. \ref{fig3} and by G. Giacomelli
{\it et\ al.} \cite{giacomelli73} in Fig. \ref{fig4}, respectively.
Thus, two approaches are possible, and we focus on Damerell's data first to find the
isoscalar amplitudes $S_{01}$, $P_{01}$ and $P_{03}$ in Eqs. (\ref{phase-1}) and (\ref{phase-2}).
We call this the set I. The other is to fit to Giacomelli's data, then, which is
called the set II.

\section{Numerical results}

\subsection{Isovector amplitude}

\begin{figure}[]
\centering \epsfig{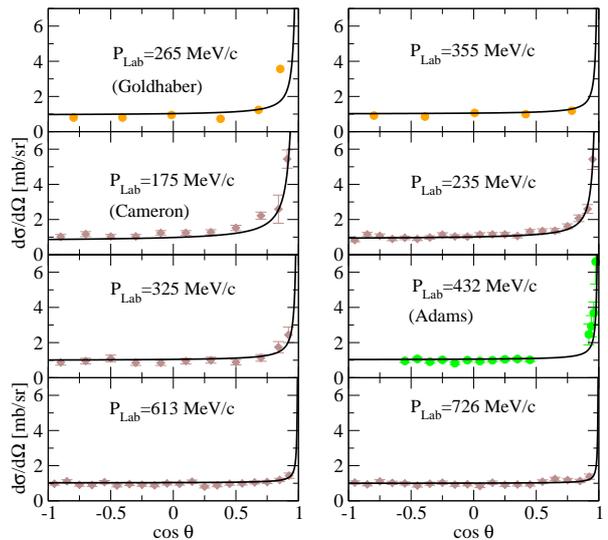}%
\caption{Differential cross sections
for elastic $K^+p\to K^+p$ reaction at low momenta. The Coulomb repulsion is
responsible for the forward peaks in the
present calculation. Data in the upper two panels are taken from Refs.
\cite{goldhaber}, \cite{cameron} and others are from Ref. \cite{adams}.} \label{fig1}
\end{figure}

\begin{figure}[]
\vspace{0.6cm}
\centering \epsfig{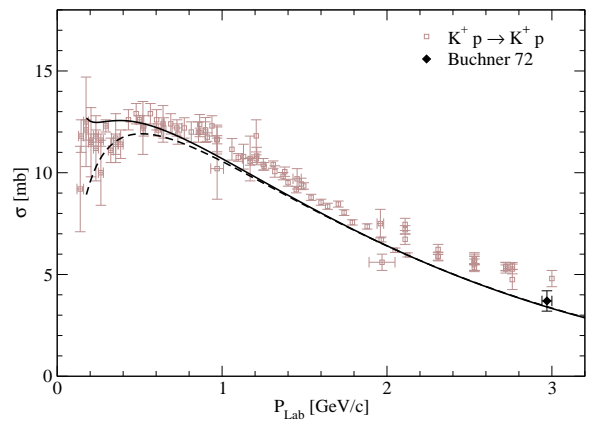}%
\caption{Total cross section for elastic $K^+p\to K^+p$ reaction
from $S_{11}$.
The solid curve includes the  Coulomb effect and the dashed one without it.
Data are taken from Particle Data Group. } \label{fig2}
\end{figure}

Given the phase shift $\delta^1_{0+}$ for the isovector amplitude $S_{11}$
in Eq. (\ref{goldhaber}), differential cross
sections $d\sigma/d\Omega$ for elastic $K^+p$ reaction in the range $150
\leq P_{\rm Lab} \leq 750$ ${\rm MeV}/c$ are shown in Fig. \ref{fig1}.
The isotropic pattern is clearly exhibited except for the Coulomb repulsion
sharply peaked at very forward angles.
Figure \ref{fig2} presents the total cross section where the solid and
dashed curves are with and without Coulomb repulsion.
Because it is highly divergent as the angle $\theta\to 0$,
we obtain the total cross section by restricting the range
of the angel to $-1<\cos\theta<0.85$ in the integration of differential
cross section.
As the s-wave with the phase shift linear in $k$ in Eq. (\ref{s11}) reproduces the
total cross section to a degree, the $s$-wave dominance assumed in the fit
is plausible up to 3 GeV/$c$, even though the anisotropy of the differential
cross section becomes stronger above the $P_{\rm Lab}\geq 800$ MeV/$c$.

\subsection{Isoscalar amplitude}

In the case of the isoscalar amplitude, however, since the anisotropy appears
in the region $P_{\rm Lab}\leq 800$ MeV/$c$ as can be seen in
Figs. \ref{fig3} and \ref{fig4},
the $p$-wave as well as the $s$-wave should be included in the
partial waves $S_{01}$, $P_{01}$ and $P_{03}$.

\begin{table}[]
\caption{Coefficients $a_i$,$b_i$, and $c_i$
for the set I and set II. The subindex $i=0,\,1,\,3$ for the isoscalar amplitudes.
The values in the left part are for the set I and the other part are for the set II.
The coefficient $b_i$ and $c_i$ are in units of GeV$^{-1}$ and GeV$^{-3}$, respectively. }
    \begin{tabular}{c|c|c|c||c|c|c}
    \hline
    \hline
     & $a_i$ & $b_i$ & $c_i $& $a_i$& $b_i$ & $c_i$ \\
    \hline\hline
    $\delta^1_{0+}$ &  3    & $-107$ & -   & 3      & $-107$ & - \\
    \hline
    $\delta^0_{0+}$ & $-36$ & 92     & 100 & $-36$  & 92     & 170\\
    $\delta^0_{1-}$ & $-5$  & $-48$  & 120 & $-460$ & $1765$ & $-2850$ \\
    $\delta^0_{1+}$ & $-12$ & 90     & -   & $-32$  & 94     & - \\
    \hline
    \end{tabular}\label{tb2}
\end{table}

\subsubsection{The set I}

The parameters for the set I are obtained from the fitting procedure to
Damarell's data \cite{damerell} in Fig. \ref{fig3}.
In contrast to the simple form of the $S_{11}$ phase shift in Eq.
(\ref{goldhaber}) the parameterization of the phase shift in
Eq. (\ref{p1}) for the $p$-wave is rather complicated due to the
inclusion of the $k^3$ term for the anisotropic angular distribution, i.e.,
\begin{eqnarray}\label{fit-1}
&&\delta^0_{0+}(k)=(a_0 +b_0 k_0+ c_0 k_0^3)\times e^{(k-k_0)/m_0}\,, \nonumber\\
&&\delta^0_{1-}(k)=(a_1+b_1 k_0+c_1 k_0^3)\times e^{(k-k_0)/m_0} \,, \nonumber\\
&&\delta^0_{1+}(k)=(a_3 + b_3 k_0)\times e^{(k-k_0)/m_0}
\end{eqnarray}
with $k_0=220$ and $m_0=100$ MeV/$c$ for $k < 220$ MeV/$c$, and
\begin{eqnarray}\label{fit-2}
&&\delta^0_{0+}(k)=a_0 +b_0 k+c_0 k^3\,, \nonumber\\
&&\delta^0_{1-}(k)=a_1+b_1 k+c_1 k^3 \,, \nonumber\\
&&\delta^0_{1+}(k)=a_3 + b_3 k
\end{eqnarray}
for  $220\leq k\leq 590$ MeV/$c$, and
\begin{eqnarray}\label{fit-3}
&&\delta^0_{0+}(k)=(a_0 +b_0 k_1+c_0 k_1^3)\times e^{-(k-k_1)/m_1}\,, \nonumber\\
&&\delta^0_{1-}(k)=(a_1+b_1 k_1+c_1 k_1^3)\times e^{-(k-k_1)/m_1} \,, \nonumber\\
&&\delta^0_{1+}(k)=(a_3 +b_3 k_1)\times e^{-(k-k_1)/m_1}
\end{eqnarray}
for $k > 590$ MeV/$c$ with $k_1=590$ and $m_1=1500$ MeV/$c$.
We use the function exponentially decreasing outside of the interval.
The continuity of the amplitude should provide a boundary condition between
two different momentum regions in order to constrain the coefficients
$a_i$, $b_i$ and $c_i$ further. They are listed in the left part of  Table \ref{tb2}.

\begin{figure}[]
\centering \epsfig{file=fig3.eps, width=0.9\hsize}%
\caption{Differential cross sections
for $K^+n\to K^0p$ (left) and  for $K^+n\to K^+n$ (right).
The solid curve results  from the set I and the dashed one from the set II,
respectively.
Data are taken from Ref. \cite{damerell}. } \label{fig3}
\vspace{0.9cm}
\centering \epsfig{file=fig4.eps, width=0.9\hsize}%
\caption{Differential cross sections for $K^+n\to K^+n$ elastic
scattering. Notations are
the same as in Fig. \ref{fig3}.
Data are taken from Ref. \cite{giacomelli73}.}\label{fig4}
\end{figure}

\subsubsection{The set II}

Giacomelli's data \cite{giacomelli73} in Fig. \ref{fig4} are used to fix
the parameters of the set II.
As before, the phase shifts $\delta^0_{0+}(k)$,
$\delta^0_{1-}(k)$ and $\delta^0_{1+}(k)$ are expressed the same as
in Eq. (\ref{fit-1}) for $k < 335$ MeV/$c$
with $k_0=335$ and $m_0=50$ MeV/$c$, and
as in Eq. (\ref{fit-2}) for  $335\leq k \leq 540$ MeV/$c$, and
as in Eq. (\ref{fit-3}) for $k > 540$ MeV/$c$ with $k_1=540$ and $m_1=3000$ MeV/$c$
with the parameters $a_i$, $b_i$ and $c_i$  listed in
the right part of Table \ref{tb2}.

Differential cross sections for elastic and charge exchange $K^+n$ interactions
are analyzed in Figs. \ref{fig3} and  \ref{fig4} based on the parameter sets I and II.
Given the different set of experimental data  by Damerell \cite{damerell}
and Giacomelli \cite{giacomelli73},
the description of the cross section from the set I is better than that
from the set II in Fig. \ref{fig3}, whereas this tendency is opposite in Fig. \ref{fig4}.
The differential cross sections at $P_{\rm Lab}=434$ and $526$ MeV/$c$ are
of particular importance to look for the $\Theta^+(1540)$ baryon,
and our fits from the set II are quite similar to those of Ref. \cite{Sibirtsev}.

\begin{figure}[]
\centering \epsfig{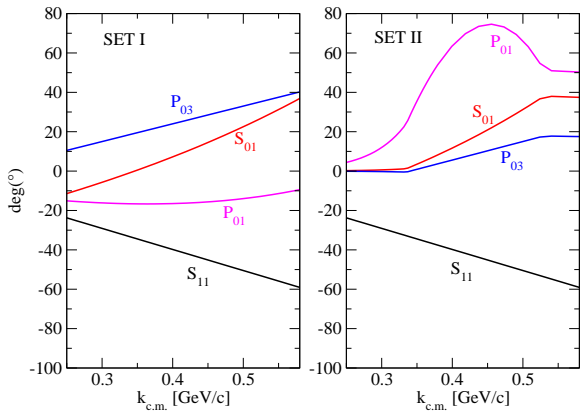}%
\caption{Phase shifts from the set I (left) and set II (right)
vs. kaon c.m. momentum $k$.} \label{fig5}
\end{figure}

Figure \ref{fig5} depicts the phase shift from the set I
fitted to Damerell's data, and from the set II fitted to Giacomelli's data, respectively.
Based on the $S_{11}$ amplitude in common, the difference is clear between
the two sets for the isoscalar amplitudes $S_{01}$, $P_{01}$ and $P_{03}$.

\begin{figure}[]
\vspace{0.6cm}
\centering \epsfig{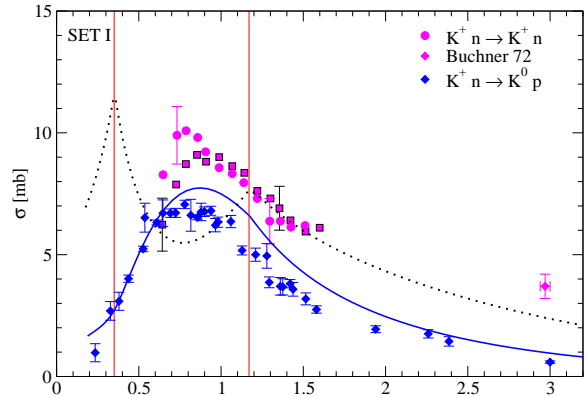}\\
\vspace{0.6cm}
\centering \epsfig{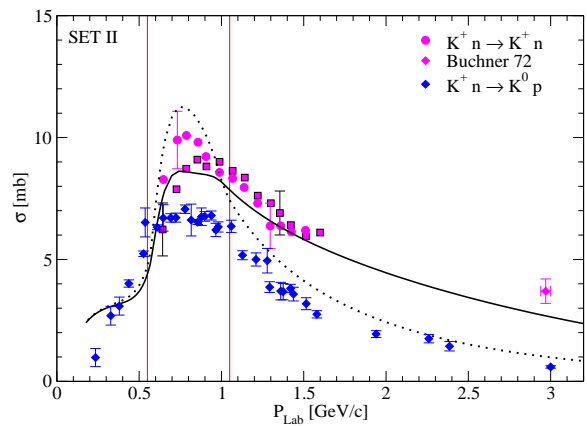}%
\caption{Total cross sections for the $K^+n\to K^0p$  and
$K^+n\to K^+n$ reactions from the set I (upper panel) and from the set II (lower panel).
The solid (dotted) curve represents the success (failure) of the parameter
set given in each panel.
Two vertical lines show divisions of the momentum range in our fits
discussed in the text.
Data are collected from Ref. \cite{giacomelli73}.} \label{fig67}
\end{figure}

In Fig. \ref{fig67} total cross sections are reproduced
for $K^+n\to K^0p$ and $K^+n\to K^+n$ reactions by using the set I in the upper
panel and  by the set II in the lower panel, respectively.
The solid and dotted curves represent the success and failure of
a given set of parameters for both reaction.
The set I agrees with the total cross section for the $K^+n\to K^0p$ reaction, whereas
the set II is consistent with the $K^+n\to K^+n$ reaction.
However, the failure of the set I for the channel $K^+n\to K^+n$ stands out,
as shown by the result of the fit convex down which is reverse to the convex up data.
The overestimate for the peak of the $K^+n\to K^0p$ reaction at $P_{\rm Lab}\approx800$
MeV/$c$ implies the disagreement of the set II with experiment either.
Thus, the two sets of parameters lead to the result contradictory to each other.


\begin{figure}[]
\bigskip{}
\bigskip{}
\centering \epsfig{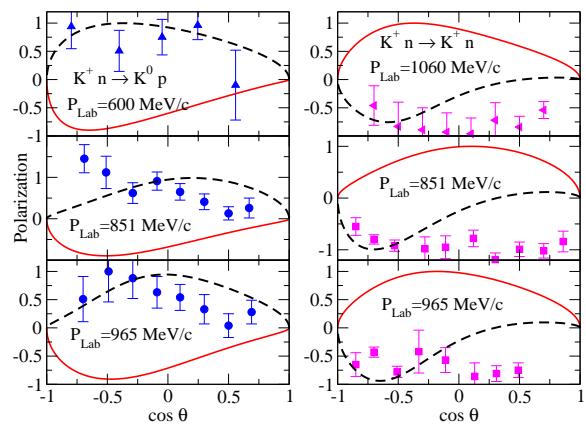}%
\caption{Polarizations for  $K^+n\to K^0p$ and $K^+n\to K^+n$ reactions.
Solid curves are from the set I, and
dashed one from the set II, respectively.
Data are taken from Refs. \cite{ray,Nakajima,Robertson,Watts}.}
\label{fig8}
\end{figure}

Summarizing what has been obtained from the set I and set II,
the total cross sections for $K^+n\to K^0p$ and $K^+n\to K^+n$ exhibit a contradiction
between the two sets of parameters, which are worse than the case
of differential cross sections. In order to find which one is appropriate for both
reactions, the polarization observable of $K^+N$ scattering is summoned for this purpose.

For the meson-baryon scattering it is given by \cite{giacomelli74}
\begin{eqnarray}\label{pol}
P={2 \textrm{Im}(f g^*)\over |f|^2+|g|^2}\ ,
\end{eqnarray}
where $f$ and $g$ are the spin non-flip and spin flip amplitudes, respectively.
Polarizations for the reactions $K^+n\to K^0p$
and $K^+n\to K^+n$ are presented in Fig. \ref{fig8} where the
solid curve results from the set I, and the dashed one from the set II, respectively.
It is interesting to note that polarizations of $K^+n\to K^0p$ are positive,
whereas they are negative in the case of $K^+n\to K^+n$. These tendencies continue
up to $P_{\rm Lab}\approx$1500 MeV/$c$ \cite{ray,Nakajima,Robertson,Watts}.
Given the sign convention for the polarization in Eq. (\ref{pol}),
it is clear that the polarization from the set II is in fair agreement
with data. The parameters for the set I
lead to the results definitely opposite to the polarization data measured
in experiments.
The sign of polarization is of significance,
because any conclusion obtained could be reversed, if the sign is reversed.
We confirm the consistency of the polarization presented in
Fig. \ref{fig8} with those experiments quoted above.
Hence, the polarization observable provides a criterion for validating
the parameters between the two sets.

\section{Discussion}

\begin{figure}[bt!]
\centering \epsfig{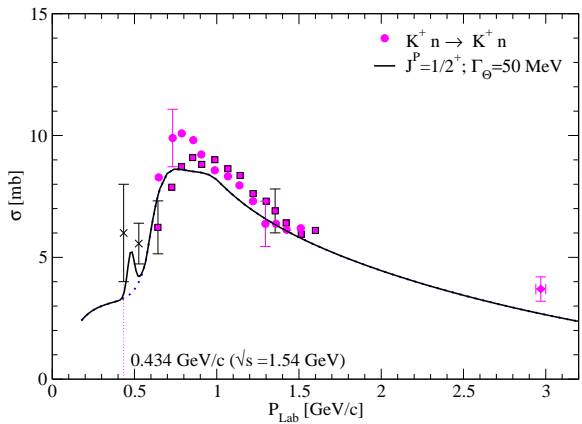}%
\caption{Total cross section for elastic $K^+n\to K^+n$ reaction.
The solid curve results from the  set II and the two points $\sigma=6\pm2$
and $5.6\pm0.8$ [mb] at $P_{\rm Lab}=434$ and 526 MeV/$c$ denoted by
the vertical bars are obtained by integrating differential
cross section data of Ref. \cite{damerell}. The $\Theta^+(1540)$ peak
is shown with parameters as discussed in the text.
Data are taken from Ref. \cite{giacomelli73}.} \label{fig9}
\end{figure}

Typically, a resonance would appear in the form of a Breit-Wigner
peak in the total cross section at the expected energy.
Therefore, the total cross section is easy to check up any profile for the
resonance peak. In Fig. \ref{fig9} total cross section for elastic $K^+n$
scattering is reproduced by using the set II.
In addition to empirical data, we introduce the
total cross section evaluated at $P_{\rm Lab}=434$ MeV/$c$ with the large error bar.
It indicates the range of total cross section $4\leq\sigma\leq 8$ [mb],
which is possible from the differential cross section data in
Fig. \ref{fig3}. With the maximum value obtained by integrating over the range
$-0.85\leq \cos\theta\leq 0.85$, the minimum is from the integration only
in the range $-0.25\leq \cos\theta\leq 0.65$
where the experimental data exist.
For further reference, a second value $4.73\leq\sigma\leq6.4$ [mb] at
526 MeV/$c$ is included in the similar fashion.
Together with the differential cross sections at $P_{\rm Lab}=434$ MeV/$c$ from the
set II in Fig. \ref{fig3}, therefore, the result in the total cross section, i.e.,
$\sigma=6\pm2$ [mb] there, strongly suggests the existence of a resonance around
$\sqrt{s}=1.54 $ GeV.

For illustration purpose we finally show the resonance peak at $P_{\rm Lab}\approx480$
MeV/$c$ which comes from the Breit-Wigner fit of Ref. \cite{ku-pn} with $J^P=1/2^+$,
$M_\Theta=1555$ MeV, $\Gamma_\Theta=50$ MeV, $I_R=\sqrt{1/2}$,  $X_R=0.25$ and the damping
parameter $d=1.5$ chosen.
A more detailed analysis of the resonance fit with these parameters could help
identifying the $\Theta^+(1540)$ baryon further, and should be pursued in
future theory and experiments.

       \section*{Acknowledgments}
This work was supported by the National Research Foundation of
Korea Grant No. NRF-2017R1A2B4010117.


\end{document}